\documentclass{article}
\usepackage{hyperref}
\usepackage{amsfonts}
\usepackage{amsmath}
\usepackage{graphicx}
\usepackage[utf8]{inputenc}

\def\bquote#1{\par\bgroup\narrower\noindent#1\par\egroup\noindent}
\def\expect#1{\langle#1\rangle}
\def\ellipsis{[\ldots]}
\def\citeH#1{[H:#1]}
\def\arXiv#1{\href{https://arxiv.org/abs/#1}{[arXiv:#1]}}

\def\Rl{{\mathbb R}}
\def\tr{{\rm tr}}

\newcounter{enotnum}
\makeatletter
 \newcommand{\linkdest}[1]{\Hy@raisedlink{\hypertarget{#1}{}}}
\def\enote#1{\expandafter\ifx\csname enot#1\endcsname\relax\refstepcounter{enotnum}\expandafter\xdef\csname enot#1\endcsname{[*\arabic{enotnum}]}\linkdest{fenot:#1}\fi%
\hyperlink{2enot:#1}{\csname enot#1\endcsname}}

\long\def\doenote#1#2\por{\par\noindent\strut\linkdest{2enot:#1}\hskip-20pt\hyperlink{fenot:#1}{\csname enot#1\endcsname}\ #2\par\vskip8pt}
\makeatother

\title{Uncertainty from Heisenberg to Today\footnote{This article is based on a talk in German, given by one of us (R.F.W.) at the 2016 annual meeting of the Heisenberg Society in Munich \cite{germanVS}. It contained some expressions of personal opinion or experience, which we left in, marked by (R.F.W.). }}
\author{Reinhard F. Werner and Terry Farrelly, \\
       Quantum Information Group, Leibniz Universit\"at Hannover}

\begin{document}

\maketitle

\begin{abstract}
 We explore the different meanings of ``quantum uncertainty'' contained in Heisenberg's seminal paper from 1927, and also some of the precise definitions that were explored later. We recount the controversy about ``Anschaulichkeit'', visualizability of the theory, which Heisenberg claims to resolve. Moreover, we consider Heisenberg's programme of operational analysis of concepts, in which he sees himself as following Einstein. Heisenberg's work is marked by the tensions between semiclassical arguments and the emerging modern quantum theory, between intuition and rigour, and between shaky arguments and overarching claims.  Nevertheless, the main message can be taken into the new quantum theory, and can be brought into the form of general theorems. They come in two kinds, not distinguished by Heisenberg. These are, on one hand, constraints on preparations, like the usual textbook uncertainty relation, and, on the other, constraints on joint measurability, including trade-offs between accuracy and disturbance.
\end{abstract}

\section*{Introduction}
Heisenberg's uncertainty relation for the momentum uncertainty $\Delta P$ and position uncertainty $\Delta Q$ of a particle,
$$\Delta P \Delta Q  \geq   \hbar /2,$$
is arguably one of the best known formulas science has ever produced.  It's almost as well-known as $E=mc^2$, but unlike Einstein's formula, there are actually jokes about the uncertainty relation, like ``No, officer, I don't know how fast I was going, but I know exactly where I am''.  What uncertainty exactly means in physics is not so easy to clarify.  An indication of this are the attempts to find a more suitable word, like indefiniteness, ignorance, indeterminacy, or imprecision.  This ambiguity already features in Heisenberg's original uncertainty paper from 1927.  Therefore, we will go back to this paper and explore these concepts. We will not do this as historians,  but rather as modern-day theoretical physicists.  That means we will consider the unhistorical question of how Heisenberg's statements look when phrased in modern quantum mechanical language.  This will be helpful for understanding the further development of ideas of uncertainty, which we will get to in the second part of this article.

\section{Heisenberg's 1927 paper}
Heisenberg submitted his paper to the journal ``Zeitschrift f{\"u}r Physik'' in March of 1927.  It was published that same year with the title ``On the intuitive content of the quantum theory of kinematics and mechanics'', or in the original German ``{\"U}ber den anschaulichen Inhalt der quantentheoretischen Kinematik und Mechanik''\footnote{\samepage We cite from this in the form \citeH{Page number}.  For the rest of Heisenberg's papers, we use the form \cite{H1}, etc., and [Number] for works by other authors.  The endnotes will be denoted by [*Number].}.

\subsection{At the threshold of quantum mechanics}
The paper appears in the middle of the explosive birth of quantum mechanics.  After 26 years of puzzles, things fell in place, and within a breathtakingly short time, the foundations of the new theory were established.  Three groups arrived almost simultaneously at formulations, which a modern physicist would recognize as versions of the current theory.  These were Born, Jordan and Heisenberg in G{\"o}ttingen, Dirac in Cambridge, and Schr{\"o}dinger in Z{\"u}rich.  The first two groups relied directly on Heisenberg's paper \cite{H0} from 1925.  Quantum mechanics then reached its full mathematical form in 1927 with Schr{\"o}dinger's work on the equivalence of the approaches, and especially von Neumann's work as an assistant for Hilbert's quantum mechanics lecture course \cite{Hil} in the Winter semester of 1926/1927.

The scientific style of the embryonic phase of quantum mechanics was strongly shaped by Niels Bohr.  The atomic model from 1913 comprised classical concepts, which were however constrained by ad hoc additional rules.  These were complemented by de Broglie's relations, connecting the particle and wave pictures. How easily one can make a misstep in this setting was exemplified by Bohr's fundamental paper itself.  What worked so well for Hydrogen \cite[Part~I]{Bo1} (except for the angular momentum of the ground state), failed completely \cite{Kra} for larger atoms \cite[Part~II]{Bo1} and molecules \cite[Part~III]{Bo1}. Likewise, the attempts of the Sommerfeld school to refine Bohr's ideas via mathematical physics soon got stuck. However, with the new breakthrough, the semi-classical style of old quantum theory became obsolete, along with particle-wave duality and Bohr's and Sommerfeld's quantization rules.  With the new quantum theory, no arbitrary rules to force the particles on a fixed path were necessary any more:\ everything that was true should now, in principle,  be justifiable from a single unified framework.

Heisenberg's first important paper \cite{H0} was written shortly before the breakthrough, while the paper \cite{H} on uncertainty followed shortly after.  Important parts are still in semi-classical style. The new theory is incorporated in the somewhat cumbersome form of ``Jordan-Dirac transformation theory''.  However, almost nothing can be seen of the mathematical language of Hilbert spaces and operators we use today. That structure was created by Johann von Neumann, and was submitted to the G{\"o}ttingen Academy in May 1927 \cite{vN}. Although both groups were in Göttingen there appears to not have been much contact between the physicists and the mathematicians \enote{1}.

\subsection{``Anschaulichkeit''}
Heisenberg's use of the word ``anschaulich'' provides a key to understanding his motivation. The German term is actually ambiguous, and Heisenberg tries to achieve his goal partly by playing with those different meanings, roughly translatable as ``visualizable'', or ``appealing to graphical imagination'' on the one hand,  and ``intuitive'' on the other \footnote{In the citations of this section all occurrences of {\it visualizable} and {\it intuitive} are actually ``anschaulich'' in the original.}\enote{transl}.
His own matrix mechanics had just been met with Schr{\"o}dinger's  competing approach. Many physicists found Schrödinger's waves, that moved around in space and time, more {\it visualizable} than the abstract matrices of Born, Jordan and Heisenberg.  The physicists of that time were well-versed in wave equations from electrodynamics and hydrodynamics, but new very little about matrices \enote{matrix}.  Wien, the professor of experimental physics in Munich, who nearly flunked Heisenberg in the doctoral exam, had told him that Schr\"odinger's work would anyhow soon supersede the atomic mysticism by Heisenberg and friends.
So providing {\it visualizable} content of matrix mechanics was important for Heisenberg to assert his priority and possibly even for the survival of matrix mechanics.  He proclaims his success in this regard near the end of the paper \citeH{196} and relegates the criticism he has thus overcome, along with the critic Schr\"odinger, to a footnote:\ ``Schr{\"o}dinger labelled quantum mechanics as a formal theory of daunting, even repulsive, {\it un-visualizability}  and abstractness'' \enote{2}.

How did Heisenberg arrive at visualizable content?  It is, after all, rather peculiar that the main new element provided in the paper, which one therefore would expect to contribute to better ``visualization'', instead imposes a {\it limitation} on particle pictures. Heisenberg's most important move is the redefinition of the term:\ away from the visualization, and towards an abstract intuition.  He begins his paper with the sentence \citeH{172}
\begin{quote}
 We believe to have understood a physical theory {\it intuitively} if we can imagine the experimental consequences of the theory qualitatively in all simple cases, and if, at the same time, we have recognized that the application of the theory will never contain internal contradictions.
\end{quote}
This quotation, which Heisenberg also used in his later years, is remarkably modern.  The term Anschauung (literally ``looking at something'') is stripped here of almost all connotations of imagining a scene or a picture.  Like the ``internal virtual images'' of Hertz \enote{3} and Galilei's geometrical figures as letters in the book of nature \enote{4}, it can just as well refer to an intuition about an algebraic or logical structure.  Whether this widening of the concept of Anschaulichkeit convinced Heisenberg's contemporary critics, however, is questionable.  Furthermore, Bohr and Heisenberg themselves were no friends of mathematically grounded intuition.  They use the terms ``abstract'', or ``formal'', or the ``symbolic character of the wave function'' \cite{Bo2} with rather negative connotations.  \enote{ansch}

An interesting feature of Heisenberg's opening sentence is the mention of ``contradictions''.  Normally, one arrives at a contradictory theory only through blunders.  From what type of contradictions should one then protect oneself?  Heisenberg's answer highlights the problematic state of quantum theory as it tries to break away from the semi-classical ``old'' quantum theory:  ``The intuitive meaning of quantum mechanics is up till now full of internal contradictions, which lead to clashes of opinions about discrete and continuum theory or particles and waves.'' \citeH{172}. Heisenberg's own work likewise pointed to such a contradiction:\  on the one hand, he  had specifically criticized the concept of trajectories of electrons in an atom in his earlier paper \cite{H0}.  On the other hand, everyone could directly see the trajectories of particles in a cloud chamber.  So how could one develop an intuition that reliably allows one to separate these two cases? This question is indeed answered in the paper.

\subsection{The microscope passage}
The famous example, via which Heisenberg develops his arguments, is simply a microscope, with which the position of an electron is determined by observation. The discussion is entirely semi-classical, i.e., the new quantum theory does not come into play.
The light allowing observation is a gamma-ray photon, which is scattered by the electron.  This interaction, known as the Compton effect, deflects the photon into an imaging instrument, through which the position of the scattering event is determined.  However, the electron experiences a kick because of this collision, so that its momentum is changed.  In Heisenberg's words,
\begin{quote}
 At the instant of the position measurement, the moment at which the photon becomes scattered, the electron's momentum changes discontinuously.  This change becomes larger when smaller wavelengths of light are used for the measurement, corresponding to more exact position measurements.  At the moment when the position of the electron is known, its momentum can only therefore be known to a magnitude, corresponding to that discrete change:\ the more exact the position is determined, the less exact the momentum is known and vice versa.  \citeH{175}
\end{quote}

It is characteristic here that the photon appears both as a wave, with a wavelength $\lambda$, and also as a corpuscle with momentum $p_1=h/\lambda$ following de Broglie, where $h$ is Planck's constant.  If we also equate the resolution of the microscope $q_1$ with the wavelength, and the change of the electron's momentum with the elastic momentum transfer, we get the relation
\begin{equation}\label{The1}
  p_1 q_1 \sim h.
\end{equation}
This is the uncertainty relation, faithfully reproduced in Heisenberg's own notation.  The tilde is (we assume intentionally) not explained by Heisenberg.  From the context, we read it as ``of the same order of magnitude''. In his own summary at the end of the paper \citeH{196} Heisenberg calls it a ``qualitative statement''.  In any case, the tilde carries the whole conceptual imprecision of the over simplified microscope theory.  With this Heisenberg again emphasizes his intention to improve heuristics, not quantitative theory.

For his simple identification of the resolution of the microscope with the wavelength of the light, first by Bohr, as recorded in the note added in proof to the paper \citeH{198}. Indeed, the mentioning  of a microscope clearly calls for Abbe's theory of microscopes, where the coordinates along and perpendicular to the optical axis play different roles and the aperture of the lens comes into play.  Furthermore, this theory can also be described in terms of semi-classical concepts and would have been able to fulfil the same function for Heisenberg's argument, while physically fleshing out the meaning of ``microscope''.  But we must defend Heisenberg from this criticism.  His argument is somewhat more abstract, and so more general, than the formulation suggests.  Instead of an electron and photon, he could have also spoken of particle $A$ and particle $B$.  The details of the optical imaging are not important \enote{5}.  Furthermore, Heisenberg could be excused here for disregarding a theory over which he almost failed his doctoral examination \cite{Rec}.

\subsection{The alleged ``proof''}\label{sec:proof}
Right after the first appearance of the uncertainty relation \citeH{173}, Heisenberg promises a proof from the commutation relations, to be given later in the paper. The only passage that fits this description is \citeH{180}, which is indeed opened with the remark that the relation \eqref{The1} can be proved by ``a slight generalization of the Dirac-Jordan formulation of quantum mechanics''. Apparently some people have taken that at face value \enote{13}. It is clear immediately that it is not a proof ``from the commutation relations'', because these do not even appear. The most commonly given modern proof \cite{Rob} does use them, thus fulfilling Heisenberg's promise to the letter. But one would be happy to have any proof from the new quantum mechanics. The ``slight generalization'' of Dirac-Jordan theory is not specified, but at least it is clear that Heisenberg is using the relation of momentum and position representation by Fourier transform, for which he cites Jordan. Clearly, this is also a sensible basis. Moreover, due to a result of von Neumann in 1931, one can derive the Fourier connection from the commutation relations.

The next strangeness is that Heisenberg identifies the accuracy of a position measurement with the width of the post-measurement state. This is clearly false in general, and we will discuss in Sect.~\ref{sec:knowledge} what may have led Heisenberg to this error. In any case he comes out of this step  with a probability amplitude ``which is only appreciably different from zero in a region of approximate size $q_1$ around $q'$'', where these parameters are accuracy (Genauigkeit) $q_1$, and measured value $q'$. The aim of the next step is to show that in that case momentum is concentrated in a region of size $p_1$ satisfying \eqref{The1}. Heisenberg discusses that this works for Gaussians, even with the tilde replaced by equality, provided we put $q_1$ and $p_1$ into the Gaussians at the appropriate places. There is no mention of standard deviations here.  But clearly this would just be an illustration, not a proof of a general fact. Heisenberg is apparently aware of this burden of proof, because he simply claims as a mathematical fact (``it will be such that\ellipsis'') that small support in position implies concentration in momentum space. Unfortunately, that statement is utter nonsense. It would imply that for amplitudes as described the uncertainty relation holds with near equality. This also shows that Heisenberg is clearly not yet thinking of the uncertainty relation as an inequality.

\subsection{What is uncertainty?}
\subsubsection{Uncertainty 1: Discontinuity}\label{sec:unc1}
The microscope argument refers to the moment of Compton scattering. Presumably, this is the moment at which the position measurement happens, so that we ``know'' the electron's position.  Precisely at this moment, the momentum of the electron changes, and we don't know exactly whether we should assign it the momentum before the interaction or that after the interaction, which has changed by $p_1$.  But  the moment of scattering is actually a poorly defined concept.  Neither in classical mechanics nor later in quantum mechanics are interactions instantaneous.  In scattering theory, the transition from one asymptote to another is idealized sometimes to a point (when a temporal course of events is even observed).  But one wouldn't normally misunderstand such discreteness as a statement about reality.  Not so for Heisenberg.  He needs the discontinuity here because it is the hallmark of ``Quantumness'' in the Bohr school just like Bohr's  quantum jumps between orbits.  More important still, the jumps are what distinguished the matrix theory from Schr{\"o}dinger's continuum theory.

\subsubsection{Uncertainty 2:\ The degree of applicability of classical concepts }\label{sec:unc2}
For the modern-day reader, an astonishing aspect of the uncertainty paper is the extent to which quantum particles are treated as classical particles.  Actually, this is intentional.  Heisenberg tries to present quantum mechanics as a minimal departure from  classical mechanics.  On one hand, this is to smooth the transition to quantum mechanics in the sense of the correspondence principle.  On the other hand, this approach helps to convince readers who are still at odds with quantum theory. For example, the equations of motion of quantum mechanics given by Heisenberg in his equation (9) \citeH{186} are identical with Hamilton's equations. At first sight this looks like a silly error, but it is actually a quote from Born and Jordan \cite{BJ}, who make a similar effort of mollifying the transition. They invent somewhat contrived definitions of the partial derivatives designed just to make this coincidence true \enote{6}.

According to Heisenberg only two small related changes to classical mechanics need to be made:\ some non-commutativity and a small precautionary measure in the form of the uncertainty relation.  These two aspects are practically identical for Heisenberg.  He treats the first as well known and the second as the new insight of his paper. On the one hand, the commutation relations offer Heisenberg a hint \citeH{173} that an uncritical use of classical concepts is problematic, and that quantum discontinuities will be felt. Conversely, his microscope analysis seems to suggest the non-commutativity:
\begin{quote}
 If, for example, the X-coordinate of the electron is no longer a ``number'', which can be experimentally inferred via equation (1), then it is the simplest conceivable assumption \ellipsis\ 
 that this X-coordinate is a diagonal part of a matrix, whose off-diagonal components express themselves in an imprecision, and express themselves in other ways under transformations.  \citeH{196}
 [Heisenberg's ``(1)'' is also (1) in this paper \enote{7}.]
\end{quote}

The difference in formalism between classical and quantum mechanics appears to lie in the commutation relations.  The difference in physical interpretation, however, is achieved by the uncertainty relation.  For Heisenberg, this parallel seems sufficient justification for viewing these aspects as two sides of the same coin and for claiming a ``direct mathematical connection'' between them.

Uncertainty relations as the limitation on applicability of classical concepts is also the explanation Heisenberg gives in his Nobel lecture \cite{Hnob}. But it is worthwhile to pause and think of what that could possibly mean. How do we ``apply'' a classical concept to a quantum particle? We can choose to think of it as a classical particle (that would be plain wrong) or as a wave (also wrong). But that is an irrelevant exercise of the imagination unless we use these concepts and make them part of the explanation of something. In that case we never use just the concept alone, but the theory context around it, equations of motion, and ideas about how to determine the various quantities. In a limited context we can compare predictions of such a classical or semiclassical prediction with quantum mechanics or with experiment, and that might be good or bad. But the ``degree of applicability of the classical concept of momentum'', as some thing that might compete with the corresponding ``conceptual position fuzziness'' is a very silly notion.

Quantum observables differ from their classical counterparts in many ways. For example, the Hamiltonian of a non-relativistic particle in a confining potential has discrete spectrum, while its classical counterpart takes a continuum of values. So if the mean level spacing $\Delta E$ is such a degree of applicability, what is the role of a conjugate $t$?. One could maybe think of something here and in other contexts, but we should renunciate the idea that there is any notion of degree of applicability of classical concepts that is meaningful in a general context or without detailed explanation.

In spite of all the efforts to make classical and quantum mechanics look alike, Heisenberg is, of course, well aware of the differences. In particular, and in spite of some statements to the contrary, it is not sufficient to just use classical mechanics with uncertainty-blurred initial conditions:
\begin{quote}
 [One could] be tempted to suppose that a ``real'' world is hidden behind the apparent statistical world, in which the causal law holds true.  But we explicitly stress that such speculations seem to us sterile and meaningless. \citeH{197}
\end{quote}

\subsubsection{The critical analysis of concepts}
The critical operational analysis of concepts is the central approach of the paper, through which Heisenberg sees himself as being in direct succession to Einstein. More specifically, Einstein's analysis and ``relativisation'' of the concept of simultaneity, with which he founded the theory of relativity in 1905 serves as a model.  Characteristic of this approach is a ``definition'' of concepts through the measurement processes.  This offers a further possibility to closely connect quantum mechanics to classical:\ in both cases, one can determine position using a microscope, which for Heisenberg now equalled a definition.  The analysis comes to a new conclusion, however, because there is a new ingredient, the de Broglie relation. This is analogous to Einstein's new conclusion about simultaneity, on account of a new principle, the source-independence of the speed of light.

In the first chapter \citeH{\S1, 174-179} of the paper, Heisenberg sketches this process, not only for position (microscope), but also for the concept of a path (a sequence of position measurements), speed (Doppler measurement), energy (collision experiments {\`a} la Franck-Hertz) and magnetic moment (Stern-Gerlach experiment).  He does not address, however, how these concepts, newly defined by a measurement process, can work together.  In classical mechanics, it is clear that one can find several measurement processes for the same concept/quantity, like a Doppler measurement or a time-of-flight measurement for momentum.  In that case, their equivalence is a result of the theory, and this provides the basis for a unified concept of momentum even in a strict operational-positivistic approach.  This is ultimately what is meant by Einstein's dictum that ``the theory determines what can be measured'' \enote{8}.  If, however, different possibilities for the definition of the new quantities are on offer, one must either choose one or clarify how these different definitions lead to the same result in the context of the new theory.

Heisenberg did not consider the difficulty of different possible ``definitions'', and already his discussion of the microscope shows that he did not mean this operational program completely seriously.  Following this program, $p_1$ would have to be the uncertainty of the electron's momentum, defined through the corresponding momentum measurement.  As a process for that measurement, Heisenberg suggests a Doppler measurement. But instead, he unceremoniously sets $p_1$ to be the momentum shift required by relativistic momentum conservation.  That there is a connection between these two uses of ``momentum'' would remain to be shown.
To put it crudely, the connection between the different variants of the ``same'' concept actually comes from ignoring the operational program.  One can hardly call this an operational analysis.

Just after citing Hamilton's classical equations of motion as the quantum ones (see above, Sect.~\ref{sec:unc2}) Heisenberg writes:\ ``The trajectory can, however, as already stated, only be calculated statistically from the initial conditions, which one can consider as a consequence of the fundamental imprecision of the initial conditions'' \citeH{186}.  Once again the concept of a trajectory appears, which was deconstructed previously as a sequence of discrete position measurements.  But it  still seems healthy enough to support a differential equation because the uncertainty of the initial conditions are invoked as the only difference with classical mechanics.  He then explicitly discusses a diffraction experiment \citeH{189}, for which classical theory gives something ``grossly different'' from the familiar diffraction image.  But ``nevertheless, we can in no way determine via the trajectory of a single electron a contradiction with the classical theory''.  This is because that ``determination'' of the trajectory would indeed require an experimental intervention, so it again hits an uncertainty barrier, and the trajectory must get destroyed.  Regarding a concept of a trajectory for unobserved particles, Heisenberg appears to have no objection, even if the clear difference between the observed diffraction image and a classical calculation already guarantees that the particles cannot have travelled on classical trajectories.

For Heisenberg there seems to be no tension between the boastful announcement of an Einstein-like conceptual analysis, and its less than half-hearted execution. Our reading is that he did not try harder because the theory he sought already existed for him - just the new quantum mechanics - and because, for this theory, the concept-critical program had not only been done already, but was even his own contribution \cite{H0}:
\begin{quote}
 Quantum mechanics had \ellipsis\ just arisen from the attempt to break with those familiar kinematic concepts and to set in their place relations between concrete values provided by experiment.  Because this appears to have worked, the mathematical scheme of quantum mechanics will require no revision.  \citeH{172}\enote{9}
\end{quote}

This link between \cite{H} and \cite{H0}, especially the application of the uncertainty idea in the new quantum mechanics, remains unfortunately rather vague.  Besides the passage quoted earlier, where non diagonal components of the matrix for position coordinate $X$ are mentioned as an expression of ``the imprecision'', there are still a few further excerpts, which we will also look at.

\subsubsection{Uncertainty 3:\ Statistics and the friction around Heisenberg's cut}\label{sec:unc3}
At the beginning of \S2, Heisenberg summarizes and ``generalizes'' his account of the concept-critical previous section, \S1:
\begin{quote}
 {\it All concepts, which are used in the classical theory to describe a mechanical system, may be also defined exactly for atomic processes analogously to the classical concepts.}\/ Experience shows, however, that the experiments, which serve such definitions, carry an imprecision if we demand the simultaneous determination of two canonically conjugate variables. \citeH{179}[First sentence emphasized in the original by additional spacing rather than italics.]\enote{10}
\end{quote}
The reference to experience is unexpected here because the arguments that Heisenberg introduces to support his uncertainty thesis are altogether purely theoretical, and are not based upon any additionally introduced experimental results.

Experiments, even thought experiments, naturally play an important role in an operational approach.  Heisenberg goes beyond this, it seems, by claiming a connection between imprecision and experiments in general.  After the remark that Born and Jordan perceive a ``characteristically statistical trait of quantum mechanics in contrast to classical theory''\citeH{177}, and that only a probability distribution can be given for variables like position, he proceeds with
\begin{quote}
 One can however, if one wishes, also say in agreement with Dirac that the statistics arise through our experiments. \citeH{177}
\end{quote}
Statistics is, on the other hand, closely connected to the uncertainty relations.  As is already stated in the abstract:
\begin{quote}
 This uncertainty is the actual reason for the appearance of statistical relations in quantum mechanics. \citeH{172}\enote{11}
\end{quote}

Hence both statistics and uncertainty happen around the act of observation.  To use a more recent terminology,  uncertainty and statistics arise at ``Heisenberg's cut'', which separates the quantum object from the instruments used for observation.  The dynamics, be it viewed semi-classically or quantum mechanically, does not include the uncertainties yet. At least that is one possible reading of a statement in Heisenberg's outlook section \citeH{197}:
\begin{quote}
 We have not assumed that quantum theory is, in contrast to the classical, an essentially statistical theory in the sense that from exactly given data only statistical conclusions can be drawn.
\end{quote}
Only the observation brings them about.  This point of view appears extensively in the examples discussed by Heisenberg and fits with his later views (e.g., \cite[46]{H3}).  It also fits with the view of two distinct quantum time evolutions:\ the deterministic Schr\"{o}dinger evolution for an isolated system and the collapse process coming with observation.

To see quantum randomness as arising in the act of observation was a very plausible idea in 1927.  This is because the observation assumes a macroscopic measuring instrument, of which one cannot possibly control all atoms with microscopic precision.  When a highly sensitive microscopic system comes in contact with a coarsely defined macroscopic instrument, random outcomes seem as inevitable as with a roll of dice.  Yet this view turned out to be completely wrong, as we will see below.

\subsubsection{Uncertainty 4:\ The degree of possible knowledge}\label{sec:knowledge}
The question of what we can know is present throughout the paper.  Already in the quote above from the microscope passage, the concern is how well the momentum is ``known''.  In his remark on Laplace's daemon, knowledge is also the central category:
\begin{quote}
 But in the precise formulation of the law of causality:\ ``If we know the present exactly, we can calculate the future'', it is not the conclusion but rather the premise that is false.  We {\it cannot} learn the present in all relevant detail, as a matter of principle. \citeH{197}
\end{quote}
The restriction of the simultaneous knowledge of canonically conjugate variables appears then like a ``data security law'' for electrons, in the form:\ ``no particle can be compelled to share his position as well as his momentum with high accuracy''.  Again, here one can ask the question of whether the particle itself knows better, or how much independence one can allow the concepts of position and momentum at the microscopic scale.

For Heisenberg, the wave function codifies the knowledge of the physicist about the system.  I (R.F.W.) remember well how puzzling I found his formulations as a student.  How should one convert something like human knowledge into a complex valued function on the configuration space?  In the uncertainty paper, Heisenberg demonstrates how he thinks this could go:\ starting with knowledge of the position with some known precision, he unceremoniously chooses a Gaussian wavepacket \citeH{180}.  But even if this choice is maybe plausible, it is hardly a guideline for the general case.

The example is from the failed proof (cf.\ Sect.~\ref{sec:proof}), and is related to its main problem:\ the false identification of the measurement imprecision $q_1$ of the microscope with the position width of the post-measurement state. This is a stronger form of the projection postulate, where not only the measured outcome but also the ``precision'' of the measurement is imprinted on the state. One may give examples where this connection holds (e.g., in typical collapse models), but just as easily one may give examples where it does not even hold approximately. But that invalidates a general proof. Heisenberg was seduced into this mistake through his concept of knowledge:\ if one has investigated the position by measurement, then one ``knows'' the position, and Heisenberg translates this knowledge in turn into a wave function.

One can avoid Heisenberg's mistake by distinguishing where the ``knowledge'' arises.  There are two fundamental possibilities for this:\ for one, we can refer to how we have made or ``prepared'' particles, and for the other, we can check properties through measurements.  In the formalism, these aspects are represented differently, namely on one hand the preparation through a density operator (or in the simplest case a wavefunction) and on the other hand the measurement through a positive-operator-valued measure (in the simplest case the spectral projectors of a self-adjoint operator).  Heisenberg does not reach this distinction and prematurely identifies knowledge from these sources. By respecting this distinction, one gets two different scenarios for quantitative uncertainty relations, which we will consider later.

In the language of quantum information theory, one could say that ``information'' is encoded in the system via the preparation, which is then read out through the measurement.  This view of quantum mechanics turned up many new questions and led to many new developments.  It would surely be too much to read Heisenberg's ``knowledge'' as ``information in the sense of Shannon'' and to make him into the founding father of quantum information theory, but, all the same, the concepts resonate to some extent.

\subsubsection{Uncertainty 5:\ Disturbance}\label{sec:disturb}
In Sect.~\ref{sec:unc1}, we had mentioned that Heisenberg does not directly address the theme of ``disturbance'' with the microscope, namely the change of the momentum by scattering, but rather emphasizes the discontinuity as an ambiguity at one point in time.  Most readers have nevertheless understood the microscope example as introducing a disturbance, and indeed the two ideas are close. Actually, the disturbance reading is also endorsed by Heisenberg himself. There is another passage (\citeH{183/184}) where the term disturbance [Störung] comes up. In this context he criticizes Jordan's notion of the interference of probabilities. He compares two time evolutions, which differ by an intermediate measurement intervention, and he identifies this change explicitly with the change of momentum from the microscope section \citeH{184 top}. And unlike in the microscope discussion bits of the new quantum theory enter the discussion. In the passage mentioned, the theme of ``uncontrollable disturbance'' appears as well, something we will later give a precise meaning to.

\subsection{Summary of reading [H]}
Many ideas simmer in this work.  They certainly do not all fit together, but in 1927 quantum mechanics was perhaps not mature enough for this to be expected.  Upon closer inspection, many of Heisenberg's arguments are not conclusive.  This was probably known to him, but it did not keep him from making bold proclamations.  It might help  to quote his characterisation of Bohr's style from the memorial volume \cite{Roz}:
\begin{quote}
 [Bohr's] insight into the structure of the theory was not the result of a mathematical analysis of its basic assumptions \ellipsis\ it was possible for him to sense the relationship intuitively rather than derive them formally. Thus I understood:\ knowledge of nature was primarily obtained in this way, and only as the next step can one succeed in \ellipsis\ subjecting it to complete rational analysis.
\end{quote}
Without a doubt, Heisenberg realised this very style in \cite{H}.  The main message, that in quantum mechanics the influence of measurements on the measured system can no longer be idealised away, is clear today and was also so clear back then that it became generally accepted immediately.

\section{What became of this}
\subsection{The Copenhagen interpretation}\label{sec:copen}
Bohr adopted the uncertainty relations practically immediately in his complementarity philosophy.  In his famous lecture in Como \cite{Bo2} in September 1927, they already played a decisive role, and they were the dominant theme in his debates with Einstein at the Solvay congress in October.  They were integrated in a new semi-classical picture of quantum mechanics:\ classical mechanics, supplemented by the uncertainty relations, and an occasional use of wave pictures.  Heisenberg also continued to write in this language in his popular accounts \cite{H1,H3,H4}.  While the difference between classical mechanics and quantum mechanics is thus understated, the public is allowed to retain a feeling of familiarity with much of the explanation.  My (R.F.W.) impression is that it is this half-hearted mix of theories that Einstein fought against in the Solvay Congress debates with Bohr, and which he referred to as ``the Heisenberg-Bohr tranquilizing philosophy - or religion?'' in a letter to Schr\"{o}dinger in 1928 \cite{Ein}.

The semi-classical language contributes much to uncertainty about the question of what exactly the ``Copenhagen interpretation'' is.  The name suggests that there should exist a manifesto from which one may learn this interpretation.  But neither proponents nor opponents of the Copenhagen interpretation have ever found such a manifesto, which could count as the interpretation of the theory on the level of 1927 (the new quantum mechanics).  Moreover, Heisenberg and Bohr were not of one mind on many important questions and also changed their minds over the course of time.  The most plausible answer to the question ``Who invented the Copenhagen Interpretation?''\ comes from Don Howard in an article with that title \cite{How}.  The answer is ``Heisenberg in the 1950s''.  Before that, there were hints of the ``Copenhagen spirit'' and of individual statements, but one cannot identify a distinct doctrine.  With his invention Heisenberg connected to the scientifically most successful time of his life and secured for himself the role of an unassailable authority on it. This was maybe especially important at that time, because his friendship with Bohr had perhaps suffered due to wartime events. In any case, his invention appealed to proponents (particularly those of the ``Shut up and Calculate'' faction) as well as opponents, who could now attack Bohr with Heisenberg quotes and vice versa.  For serious discussions about foundations of quantum mechanics, it is advisable to avoid the term completely.

One of the most coherent texts regarding the Copenhagen interpretation is the chapter with this title in Heisenberg's book \cite{H3}.  This is noteworthy in our context because of a curious appearance of the uncertainty relation:\ according to Heisenberg, it has to be taken into account when describing an experiment, particularly for the ``translation of the initial experimental situation into a probability function''. However, uncertainty is the least concern during this translation because even the most ineptly chosen wavefunction automatically fulfils the uncertainty relation.  But this remark offers Heisenberg the opportunity to once more bring his big discovery of 1927 into play\enote{14}.

The preference for semi-classical thinking could have also contributed to the fact that the Copenhagen school missed entanglement, which, according to Schr\"{o}dinger, was the essential new trait of quantum mechanics \enote{15}.  The situation of two distant non-interacting but correlated systems became foundational for quantum information theory and of course goes back to the paper of Einstein, Podolsky and Rosen (EPR)\cite{EPR}.  The idea is to replace a disturbing observation on one system by a ``remote sensing'' measurement on the other, given a suitably correlated state.  Several  of the old puzzles of quantum mechanics are affected by this.  To begin with, this consideration completely destroys the view (see Sect.~\ref{sec:unc3}) that the quantum mechanical randomness could be traced back to the interaction with coarsely specified macroscopic measurement apparatuses (see end of Sect.~\ref{sec:unc3}).  This is because, for every measurement on a subsystem, there is a ``partner measurement'' on a distant system, so that the measurement value pairs perfectly agree.  If one only observes one system, one sees the usual quantum mechanical randomness.  But the idea that the ``insufficient specifications'' of the two measurement apparatuses bring about the randomness, while the results are perfectly correlated over a large distance, is absurd.

Important consequences of the EPR example arise also for the description of subsystems if one assumes a minimum of locality/signal causality.  The non-existence of joint measurements also follows, which is an important foundational element of complementarity.  One might think that such thoughts would have perfectly suited the Copenhageners, or could have been a stimulus for further development.  But instead the EPR paper was seen as an attack and was ``parried''.  The background for this was  the enormous success of quantum mechanics, with which many young and brilliant researchers were just making their careers.  To them the misgivings of the old man Einstein were just annoying, and had to be neutralized quickly in order to return to work.  Bohr was the ideal authority to do that, and probably few people bothered to read his reply. It did not matter that he missed all the new and interesting points about separated laboratories and entanglement, by transforming away these aspect in a footnote. He mainly embarked on a sermon criticizing the EPR authors for their lack of conformity with his complementarity ideas, so everything was back to normal.

Nevertheless, Bohr's magic had the desired effect.  Peter Mittelstaedt, who did his doctoral thesis and habilitation with Heisenberg, was in the audience of one of the various lectures that I (R.F.W.) gave on the subject.  He confirmed my interpretation that \cite{EPR} was mainly perceived as a nuisance.  When he told Heisenberg that he wanted to work on the ``EPR-paradox'', as it was called then,  Heisenberg answered ``What do you want with that then; Bohr has dealt with that for us''.

One difficulty with the Copenhagen texts is that they do not necessarily strive for clarity.  Heisenberg as well as Bohr liked to quote the aphorism that the opposite of a deep truth is also a deep truth. Or even worse, that clarity and truth are complementary.  Clarifying something often requires a decision between conceivable conceptual alternatives, but that would have reduced the profoundness for them.  When H.P. Stapp made the attempt in 1971 to give a precise formulation of the Copenhagen interpretation, he asked Heisenberg for a confirmation of his version.  He gave it (for us astonishingly), but added a general criticism of this intention:
\begin{quote}
 It may be a point in the Copenhagen interpretation that its language has a certain degree of vagueness, and I doubt whether it can become clearer by avoiding this vagueness. \cite[1113]{Sta}
\end{quote}
Thus, no one can be certain to have found the Copenhagen Interpretation.  What remaining value it has can only be very subjectively judged.  Often enough I (R.F.W.) have been called a ``Copenhagener'', and actually there are some often cited fundamental views, with which I would agree.  These include the necessity to capture experiments and their results in classical language.  However, it makes a huge difference whether one applies these statements to classical properties of the measurement apparatus, or - much more problematically - to properties of the microsystems themselves.  As for complementarity, it is certainly true that one must always make a choice, when running an experiment or theoretically describing it,  and different choices preclude each other.  But Bohr's love for contrasting pairs of concepts can safely be ignored. Orwellian doublethink, i.e., the feat of holding two contradictory beliefs simultaneously, like a particle and a wave picture, has no place in quantum mechanics.
Personally, I sympathize with the pragmatic stance of avoiding ontological debates.  But perhaps many would already disagree with me here.

\subsection{Minimal statistical interpretation}
For the further discussion, we would like to rely on an interpretation of quantum mechanics, which radically implements Heisenberg's program of referring only to quantities that can be measured.  Namely, the ``quantities'' are not to be understood as properties of particles, but rather are defined through a measurement process.  Here ``process'' is to be understood entirely in the sense of laboratory language, meaning it refers to apparatuses, which are to be thought of as being described in the language of classical physics.  The counterpart of such apparatuses in the theory is the specification with which probability the different possible measurement values appear, when the apparatus works on a given preparation/state.  This assignment is called an observable.  Position and momentum are observables in this sense, but it is never assumed that single particles have a known or unknown value of these quantities assigned to them (either at a particular time or always).  For example, the
position observable is essentially defined by Born's rule making $|\psi(x)|^2$ the probability density for finding the particle at $x$. In the further development of the theory the observables may play different roles, for example regarding symmetries, or for writing down interactions. But none of this makes the values of observables more like properties of individual systems. Similarly, the states are not defined through particle properties or a distribution thereof, but rather through the preparation process.  The statements of the theory all concern  probabilities that can be measured by macroscopically described experiments.  We call this interpretation the minimal statistical interpretation.

It was shown by the axiomatic program of G\"{u}nther Ludwig \cite{Lud} that one can build the usual quantum mechanics upon this. For my current primary research area, quantum information theory, this interpretation likewise offers an adequate basis. Interestingly, this interpretation is by no means the majority view in that field. For example, David Deutsch constructed the first quantum algorithms as evidence for the (not at all minimal) many-worlds interpretation of quantum mechanics. But we can easily agree with Deutsch on whether a proposed algorithm ``works'' because this can be formulated and decided on the basis of the minimal interpretation.  In a community of researchers with wildly different ideas about foundations, the minimal interpretation  defines the common ground.

The most interesting question is then if and how one can go beyond the minimal platform and come closer to an ``objective'' description of quantum systems.  Naturally, that was often tried and will certainly be tried again.  But the grander the attempt the more it typically fails. What always works is to come to an objective language in a limited context, which might even be justified from an approximation of quantum mechanics.  For example, a time-of-flight spectrometer is treated like a device filtering classical particles, we talk of particles ``going through'' slits, or discuss the formation of atoms by placing electrons into shells (a metaphor based on the Hartree-Fock approximation). Further examples  are the classical Maxwell theory for all quantum optics experiments, in which photon-photon correlations are not considered, or even semi-classical mechanics if one restricts oneself to observables that change slowly in phases space on a scale given by $\hbar$ \cite{WW,Wcl}. The language one actually uses in laboratories is full of such classical elements, and how could it not be? It is natural language after all, which has developed as a way to communicate about perfectly classical, property-defined things. We have an innate tendency to tell stories, and to base ``understanding'' on moment-by-moment accounts of what is happening. The quantum mechanical formal language is completely frustrating in this regard. It never tells us what is happening, but only what we will find (and how often) if we look to find out.

What is the relation of this lab talk to the theory then? Of course, we need it whenever we want to get quantitative agreement. But it is also needed around the boundaries of applicability of the classical elements. For example, the Hartree-Fock approximation is by far not the best we have, and unless one wants to hamper progress, one has to get used to the idea that in a multiparticle electron wave function it makes little sense to ask which one-particle states are ``occupied''. Even more importantly, the partly classical lab talk, like Heisenberg's semiclassical theory, is not free from contradictions and paradox. Then it is important to distinguish whether there is really something wrong, or if once again our classical cognitive reflexes have led us astray. There is actually a whole literature of Gedanken experiments, in which one raises classical expectations, only to find them shattered, and then to marvel at the quantum strangeness.

And so we come back to Heisenberg's programmatic statements about the heuristic meaning of the uncertainty relations.  Only now the logic is turned around:\ we do not need semi-classical arguments to justify the uncertainty relations,  but rather the relations, now as general theorems of quantum mechanics,  help to identify the regime in which semi-classical arguments are justified.

\subsection{Making the uncertainty relation precise}
\subsubsection{Uncertainty 6:\ Preparation uncertainty}
Refining the heuristic uncertainty idea and developing it into general statements about the (new) quantum mechanics began already in 1927.  Earl H.\ Kennard came on a sabbatical year to G\"{o}ttingen and probably followed the young Heisenberg to Copenhagen because his work  \cite{Ken} ``On the quantum mechanics of simple types of motion'' gave Copenhagen as his address.  Also, for correcting the galley proofs, he apparently met Heisenberg in Munich \cite[p.588]{Rec}.  In this clearly written paper, we find the uncertainty relation, as it appears in every textbook on quantum mechanics today.

Like Heisenberg, Kennard relates uncertainty to ``knowledge'', but of the special kind that we get from having prepared the system.  His relation says that it is impossible to prepare particles that have both their position probability distribution and their momentum probability distribution sharply concentrated around a single value.  As a precise quantification of sharpness of the distribution, he uses the standard deviation $\Delta A$.  That is, for a general observable $A$ with real values, he has
\begin{equation}\label{stdev}
\Delta A^2=\expect{A^2}- \expect A^2,
\end{equation}
where the brackets represent the expectation of the quantities in the given state.  His proof is somewhat awkward.  The currently used proof follows Robertson \cite{Rob} in the use of the canonical commutators, fulfilling Heisenberg's unfulfilled promise of such a derivation.  Independently from Kennard, in 1927/28 Weyl also arrived at this relation in his book \cite{Wey}, though he did not cite Heisenberg and thanked Pauli for the tip about the uncertainty idea.

One can hardly overestimate the advance from Heisenberg to Kennard.  In contrast to a relationship demonstrated through example in the semi-classical old quantum mechanics Kennard provides a general theorem in the new.  Instead of Heisenberg's vague tildes he gives a mathematically sharp, quantitatively falsifiable relation.  And finally, the vast ambiguity of the interpretation is reduced to zero.  Anyone who wants to know what Kennard's uncertainty relation means, only needs to check what the proof actually proves.  So it is not astonishing that this relation (with simplified proof) became ``the'' uncertainty relation in the pedagogical literature.  Whenever students today are asked about the precise meaning of the Heisenberg uncertainty relation, they are supposed to reproduce those parts that did not come from Heisenberg.  Heisenberg himself saw this differently.  In his lectures in Chicago, he gave Kennard's proof and then deprecated Kennard's achievement with the words
\begin{quote}
 It should still be emphasized that this derivation is, in its mathematical content, in no way different from the derivation of the uncertainty relation from the wave-particle duality; only the proof is \ellipsis\ conducted precisely here. \cite[Sect.~II.1.]{H1}.
\end{quote}
What?  We read that as ``a proof adds nothing to the mathematical content (!), and I share the glory for this discovery with no one'' (see also \enote{9}).  It would be interesting to discover Kennard's reaction to this assessment and to see whether it was this appreciation that made him change fields \enote{16}.  The further development of uncertainty relations practically ground to a halt for a long time due to the success of Kennard's formulation.  The relation of Robertson \cite{Rob} is often called a generalisation to arbitrary observables, but actually it is not an uncertainty relation in the sense discussed here, because the lower bound is still dependent on the state and, for eigenstates of one of the observables, it tells us nothing.  It does not allow the conclusion that ``the distributions for the same state cannot both be sharp'', not even in cases where this statement is actually true.  That leads to the challenge to set up and prove uncertainty relations for other observables.

\begin{figure}
  \centering
  \includegraphics[width=5cm]{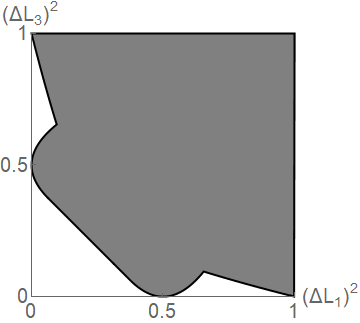} \includegraphics[width=5cm]{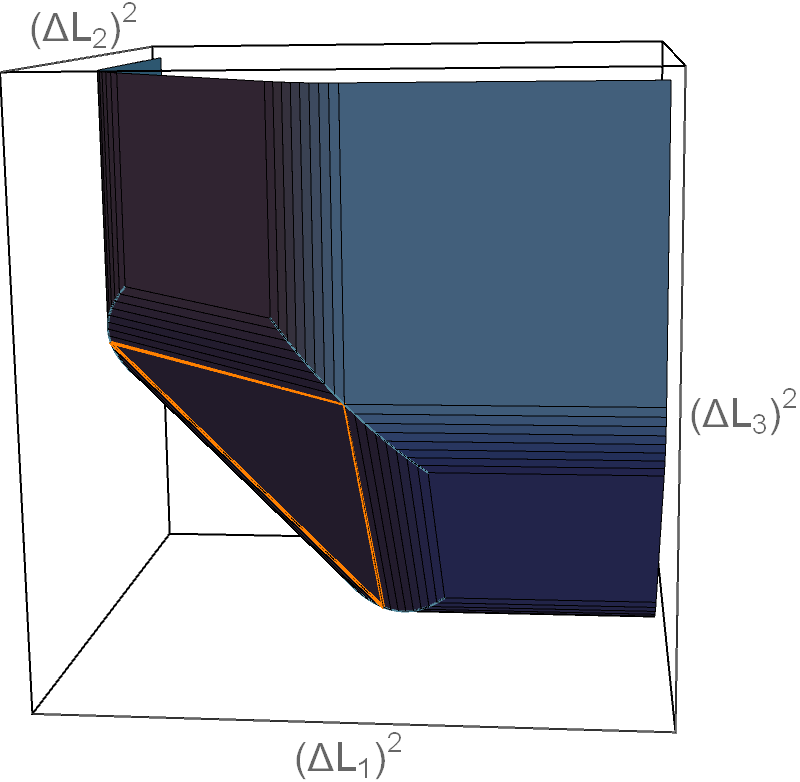}
  \caption{Uncertainty regions. Left:\ for two components of angular momentum of a spin-1 system. Due to the concavity of variance this is not a convex body.
           Right: for all three components, with the additional simplification that we included with every point all those where one or more variances is larger. This is the relevant body if one is interested in lower bounds on uncertainty only. For details and a paper cut-out model, see \cite{DSW}}\label{fig:1}
\end{figure}

In the simplest setting, one chooses the variance $\Delta A^2$ for the ``width of a probability distribution'' as Kennard did.  A general uncertainty relation for two observables $A$ and $B$ is best given in the form of a diagram (see Fig.~\ref{fig:1},Left) that shows the possible values $(\Delta A^2,\Delta B^2)$, taken in the same state, and collected for all states.  An uncertainty relation is then every inequality that allows the conclusion that the point $(0,0)$ cannot be reached.  There are simple algorithms \cite{DSW,SDW} to calculate the boundary curve, which unfortunately cannot be calculated analytically in most cases.  Sometimes it is also natural to consider more than two observables, like, e.g., the three components of angular momentum (see Fig.~\ref{fig:1},Right).  A further interesting development arises if the ``sharpness of the distribution'' is represented in an information theoretic sense via entropies \cite{CBTW}.

\subsubsection{Uncertainty 7:\ Measurement precision and disturbance}
Most people have read Heisenberg's relation as a trade-off between the precision $\Delta Q$ of an approximate position measurement and the momentum disturbance $\Delta P$ incurred by that measurement.  This has nothing to do with the preparation uncertainty because it refers to a completely different experimental situation.  To verify a preparation uncertainty relation, one has to perform separate experiments for every state and every relevant observable (e.g., $P$ and $Q$) to record the distribution.  No single particle is thereby subjected to both a position and a momentum measurement.  On the other hand, this is essential for any precision-disturbance trade-off.

A common approach to this distinction is to ignore it, namely first to prove the Kennard relation and then to pretend that this says something about measurement precision and disturbance.  One finds this in many textbooks and, in a way, this follows Heisenberg, who did not make this distinction either \enote{17}.  It is, however, pedagogically unfortunate to prove something and then lie about the conclusion.  A more honest approach is to at least discuss an example of disturbance (e.g.,  \cite[Chap.~IV.III]{Mes}). But can one express this idea as sharply and as generally as for preparation uncertainty?

It was already known to Kennard that his new relations did not cover the categories of precision and disturbance.  He remarks explicitly that a definition of ``measurement error'' as a deviation of the measurement value from the true value does not work in quantum mechanics because the true value does not exist ``in a physical sense''.  He recommends a comparison of the probability distributions.  He does not take this further, but we will follow this line of thought here.

In the minimal statistical interpretation, observables are thought of as describing a measurement process.  If one changes the measurement setup, there is no point in discussing what would have come out for the original one (no counterfactual definiteness, or ``unperformed measurements have no results''). Therefore, comparing the momentum after the microscope measurement with the value one would have obtained without the microscope makes no sense in the theory. Likewise, a single run of the microscope experiment does not produce a value of ``momentum disturbance'' of which we could then collect the statistics by repetition. Of course, it is also not an option to measure the momentum before the microscope:\ that would surely disturb the position measurement, and we cannot ignore this effect, since this kind of disturbance is precisely the point under discussion.

On the other hand, two measuring devices are described by  the same observable, if they give the same statistics on all states. It is natural to replace ``the same'' here by ``almost the same'', i.e., to compare observables not by individual results but by comparing {\it distributions}. This way of detecting a disturbance is also familiar from other contexts.  Consider, for example, the double slit experiment.  It is well-known that if one tries to find out through which slit the particles go by detecting their passage at the slits, the interference pattern will be destroyed. That is, the particles get disturbed. This statement does not require a dubious comparison of an actual trajectory with a hypothetical undisturbed one. The change of the distribution on the screen is enough.

Hence in order to identify a disturbance in the microscope experiment we can compare the momentum distribution of the particles after the measurement with that obtained by a direct momentum measurement. From this comparison, made for all input states we get $\Delta P$. The same idea applies to accuracy:\ we compare the output distributions from the given device with those from the observable one intended to measure (the ideal, or ``reference'' observable). For the microscope this standard of comparison is the position observable, which is implicit by calling the microscope an ``approximate position measurement''. The accuracy is a property of the device, a benchmark quantity. Some value $\Delta Q$ implies the promise that, no matter which input state is chosen, the deviation between probability distributions from the device and the reference is less than $\Delta Q$.

What is conspicuously absent from the above explanations, but is clearly needed to come to quantitative relations, is a way of quantifying the deviation between two probability distributions. There are different ways of defining this, which we will briefly describe in Sect.~\ref{sec:compareP}. But once this is settled, one may embark on finding quantitative trade-off relations between $\Delta P$ and $\Delta Q$. Moreover, it is clear that there is nothing special about position and momentum in this conceptual framework, so it applies to arbitrary pairs.

\subsubsection{Uncertainty 8:\ Measurement precision and uncontrollable disturbance}
Heisenberg \citeH{183} also refers to the disturbance from a measurement procedure as ``fundamentally uncontrollable'' (e.g., also  \cite{Mes}).  For a long time, I (R.F.W.) found this expression peculiar, but I believe we can now give a good explanation.  This is based on the question of how one could  try to {\it control} the disturbance.  Let us assume from the outset that we know the construction of the microscope exactly.  With that we know all systematic errors, which one can possibly correct for.  The measurement after the interaction  then no longer needs to be the standard momentum measurement because it may contain such corrections. We can allow an essentially arbitrary measurement device, constructed with the sole purpose of producing momentum-like outputs, whose distribution is always as close as possible to that of a direct momentum measurement. We can even grant this reconstruction measurement access to the position value obtained in the prior position measurement, or to internal quantities obtained from some monitoring of the microscope measurement, as long as we do not interfere with the process of determining $Q$, i.e., as long as the microscope is not ``disturbed'' in the sense explained above. An uncertainty relation for precision and ``uncontrollable'' disturbance is then a trade-off between $\Delta Q$ and $\Delta P$, obtained with an optimized reconstruction measurement.

\subsubsection{Uncertainty 9:\ Measurement uncertainty}
We can generalize this further and at the same time restore the symmetry between position and momentum.  Taking the microscope and the reconstruction measurement together as one big device, we will consider a scenario in which there is only one device with two outputs, one position-like and the other momentum-like \enote{povm}. We can then determine $\Delta Q$ as before, by evaluating the difference between the position-like output distribution with that of an ideal position observable. In this process the momentum-like output is ignored. Our previous definition of momentum disturbance still applies, only this time we ignore the position-like output. That gives a mirror image of the accuracy definition, so in this scenario disturbance is actually the same as accuracy for the momentum-like output.
For position and momentum, there are measurement uncertainty relations of the form
$$\Delta P \Delta Q \geq a \hbar ,$$
where the form of the relation already follows from dimensional analysis, namely from the symmetry $Q\mapsto\lambda Q$ and $P\mapsto P/\lambda$.  The constant $a$ depends still on how we compare distributions, as also occurs for the corresponding constant in the preparation uncertainty relation.  The first relation of this form is proved in \cite{We1}.  Further developments are found in \cite{BLW1,BLW2}, and some are described in Sect.~\ref{sec:further}.

\subsubsection{Technical supplement:\ Quantitative comparison of probability distributions}\label{sec:compareP}
The crucial technical point for the definitions given above is the evaluation of the distance between two probability distributions, which plays the same foundational role for measurement uncertainty as the variance plays for the evaluation of the width of a distribution for preparation uncertainty.

In some applications, the ``total variation'' is used as the distance between two probability measures.  Up to a factor this is the largest difference between the probabilities of any event if first one and then the other distribution is used in the calculation.  Such a distance measure has the physical unit of a probability and so is a dimensionless number.  However, we want a metric that reflects distances in the underlying outcome space. So the distance between two distributions of position should be in meters, and the distance between two sharply peaked distributions should be roughly the distance of the peaks.

The general approach starts from this consideration and also works for measurements with arbitrary outcomes, not necessarily real numbers.  We take the outcomes as elements of a set $X$, on which we have already defined a metric. That is, the main input to the construction is a way to ascertain the distance $d(x,y)$ between two points $x$ and $y$.  For the real-valued case, one usually sets $d(x,y)=|x-y|$.  For other quantities, for example, for angles \cite{BKW}, there are many natural choices depending on the application at hand.  We denote by $\delta_x$ the probability distribution that delivers the value $x$ with certainty.  Mathematically we call this a point measure, and the $\delta$ is reminiscent of Dirac's $\delta$ function.  Then it should be true that
$$d(\delta_x, \delta_y)=d(x,y).$$
Here we have simply taken the same letter ``$d$'' for the distance between measured values and for the (not yet defined) distance between distributions to emphasize the connection between them, which is expressed through this equation.  As a next step, we consider the distance between a point measure $\delta_x$ and a general distribution $\mu$.  The distance of a measured value $y$ from $x$ is then a random variable $d(x,y)$.  More generally, we consider the power $d(x,y)^\alpha$, with a so-called error exponent $\alpha \geq 1$.  For the usual variance and related quantities, we take $\alpha =2$.  We then choose the distance to be the $\mu$-expectation value of this distance function (with power).  Following the usual rules of probability theory, this is given by an integral over $\mu$:
$$d(\delta_x, \mu )^\alpha  = \int\mu (dy)  d(x,y)^\alpha   = \expect{d(x,\cdot)^\alpha}_\mu    $$
Here the power $\alpha$ on the left side of the equation ensures that $d(\delta_x, \mu )$ has the same scaling and units as the metric $d$ itself.  In this framework, the natural spread of a probability measure $\mu$ is
$$\Delta (\mu )=\min_x d(\delta_x, \mu ) , $$
i.e., the distance to the nearest point measure. One easily checks that for $\alpha=2$ and $(X,d)=(\Rl,|\cdot|)$ this is just the standard deviation \eqref{stdev}.

For the distance between two arbitrary probability measures $\mu$ and $\nu$, one could now think about averaging the metric in two variables.  But that is a bad choice because then $d(\mu ,\mu )$ would rarely be zero.  Instead, one averages over a joint distribution for the two measures, i.e.,  a measure $\gamma$ on $X\times X$, so that integrating out the second variable gives $\mu$ and integrating out the first gives $\nu$.  Such a measure is called a ``coupling'' of $\mu$ and $\nu$ or a ``transport plan''.  This goes back to a task tackled by the mathematician and fortress builder Gaspard Monge, in which heaps of clay with distribution $\mu$ must be converted into a fort with clay distribution $\nu$.  The measure $\gamma$ then describes how much earth should be moved from the vicinity of $x$ to the vicinity of $y$.  We now assume that the transport of a bucket of earth from $x$ to $y$ incurs a cost of $d(x,y)^\alpha$.  For example, with $\alpha=1$ this means paying workers by the bucket and the meter. With $\alpha=2$ they get a bonus for long hauls. Then the cost of moving earth for building the fort, with a cleverly chosen transport plan is
\begin{equation}\label{distprob}
  d(\mu ,\nu )^\alpha = \min_\gamma  \int\gamma (dx dy)  d(x,y)^\alpha .
\end{equation}
One therefore calls $d(\mu ,\nu )$ the {\it transport distance} from $\mu$ to $\nu$.
This is indeed a metric on the set of probability measures with finite variance and has many desirable properties.  A good book on the subject is \cite{V}.

When considering observables we must now take into consideration that every state $\rho$ that is measured gives another probability distribution.  We denote by $\rho_A$ the probability distribution that arises via the measurement of an observable $A$ on the state $\rho$.  Here, $A'$ can also be a generalized observable (POVM, see \enote{povm}) as is typical of the marginals of joint measurements.
\begin{equation}\label{obsdist}
  d(A',A)  =  \max_\rho  d(\rho_A, \rho_{A'}).
\end{equation}
In the typical uncertainty application $A$ might be the observable that we want to measure, the ``ideal reference'', e.g., the standard position observable. $A'$ will be an approximate version of it, for example one marginal of a joint measurement.  The distance $d(A',A)$ is then an overall figure of merit for $A'$, that one might find in the specs of the device $A'$.  The specification $d(A',A)\leq\varepsilon$ is equivalent to the following promise:\ regardless of which state $\rho$ is measured, the distribution $\rho_{A'}$ arising from measuring $A'$ deviates by at most $\varepsilon$ (in the sense of transport distance) from the distribution $\rho_A$, which the reference observable $A$ would have given.

It is important that this holds for all states, i.e., that the maximum is taken in \eqref{obsdist}. The state-dependent quantity $d(\rho_A, \rho_{A'})$ by itself would be a ridiculously weak measure of quality. A testing lab using it could be fooled by a device $A'$ which simply outputs the distribution $\rho_A$ on every state. Of course, this hardly deserves being called a measurement. A ``good measurement'' should be one that delivers reliable results even on unknown states.

\subsubsection{Results on measurement and preparation uncertainty}\label{sec:further}
The definitions given above allow us now a quantitative specification of measurement uncertainty relations.  As in the preparation uncertainty case it is a good idea to draw an uncertainty region, i.e., the set of pairs
$\bigl(d(A',A), d(B',B)\bigr)$, where $A',B'$ are the marginals of some joint measuring device. A measurement uncertainty relation is any statement to the effect that this set of pairs does not reach the origin. Apart from the observables $A$ and $B$ under consideration this region depends on all the choices made for the quantitative description of uncertainties:\ the metrics on the outcome spaces and the error exponents, which need not be the same. Note that these choices determine a variance quantity \eqref{stdev} for use in a preparation uncertainty relation as well as a distance \eqref{obsdist} for probability distributions and observables. Therefore, it makes sense quite generally to compare the preparation uncertainty diagram with the measurement uncertainty diagram.

Doing this for position and momentum we find that the diagrams are exactly equal! Taking Euclidean distance and $\alpha=2$ this gives \cite{BLW1,BLW2}
\begin{equation}\label{meas2}
 d(Q',Q) d(P',P) \geq  \frac\hbar2\ .
\end{equation}
with the same constant as in the Kennard-Weyl preparation uncertainty. To understand what is going on, let us consider a more general case, namely that of two observables related by Fourier transform. The outcome sets of these observables can be arbitrary locally compact abelian groups, which are duals of each other. The product of these spaces is then called a phase space. The Hilbert space is the space of square integrable functions on one of these groups with Haar measure, and it does not matter which, because in this general setting we have a unitary Fourier transform connecting the $P$ representation and the $Q$-representation. In the standard position/momentum case both groups are $\Rl$, but we could also take $\Rl^n$ if we think of position and momentum as vector valued, or the circle group and the integers (angle and number)\cite{BKW}, or bit strings, or combinations thereof. Of the two metrics we only demand translation invariance $d(x+z,y+z)=d(x,y)$, which makes sense because the domain is a group. We claim \cite{We3} that in all these cases the uncertainty regions for measurement and preparation coincide. The crucial step is to consider a special type of joint measurement, called a {\it covariant phase space measurement}. These have the property that phase space shifts on the input quantum state are equivalent to the corresponding shifts of the outcome distributions. The structure of such observables is known. Each observable is uniquely characterized by an operator $\sigma$, which gives the probability density at the phase space origin. This is sufficient to determine the observable, because by covariance we then get the density at all phase space points. The necessary and sufficient condition for $\sigma$  to be the density of a normalized covariant observable is that $\sigma\geq0$ and $\tr\sigma=1$, i.e., $\sigma$ is a density operator as is normally used to describe a preparation. These joint measurements are well-known in quantum optics, where taking $\sigma$ as the oscillator ground state as a density gives an observable whose output density is the so-called Husimi function of the input state.

One can compute the marginals of such an observable, and it turns out that the $Q$-marginal is the convolution $\rho_{Q'}=\rho_Q\ast\sigma_Q$, where these are the output densities for $\rho$ and $\sigma$ for $Q$, and similarly for $P$. In other words, we can simulate the marginal by making a standard position measurement on $\rho$ and then adding noise with distribution $\sigma_Q$ from an independent source. This immediately provides the idea for a coupling $\gamma$ of the distributions $\rho_Q$ and $\rho_{Q'}$, namely the measure on $X\times X$ where the first component $x$ is distributed according to $\rho_Q$ and the second is $y=x+z$ with $z$ distributed according to $\sigma_Q$. Doing the integral in \eqref{distprob} then gives
\begin{eqnarray}\label{distconv}
  d(Q',Q)^\alpha
     &\leq&\sup_\rho\int\rho_Q(dx)\sigma_Q(dz)\ d(x,x+z)^\alpha \nonumber\\
     &=&\sup_\rho\int\rho_Q(dx)\sigma_Q(dz)\ d(0,z)^\alpha \nonumber\\
     &=& d(\sigma_Q,\delta_0).
\end{eqnarray}
That is, the errors are bounded by the size of the added ``noise'', which in turn is a variance, once we shift $\sigma$ in phase space so that the minimum in the variance definition is attained at $0$. So the measurement and preparation uncertainty diagrams are equal by virtue of a one-to-one correspondence between (covariant) joint measurements and states. Via this mapping we can also take over the minimizers and conclude that the unique covariant phase space measurement for which \eqref{meas2} is equality is the one giving the Husimi distribution. What is left for a general proof is an argument for why non-covariant joint measurements cannot do better. This is done by an averaging argument, but also involves a careful discussion excluding that some averaged observable picks up some non-zero probability for infinite values \cite{BLW2}.

From this proof sketch it is clear that the quantitative equality of preparation and measurement uncertainty is due to the high symmetry of the observables connected by Fourier transform.  If one were to choose two arbitrary observables $A$ and $B$, such a relation rarely holds.  Indeed, there are efficient ways to compute measurement uncertainty bounds by semidefinite programs \cite{SRW2}, and even more efficient ways \cite{SDW} for preparation uncertainty mentioned above, so it is easy to generate examples. This makes it clear that Heisenberg's identification of the two kinds of knowledge is indeed invalid.

For a direct example of this kind consider for each observable $A$ and $B$ a complete von Neumann measurement, i.e., a projective measurements along some orthonormal basis.  Then the uncertainty pair $(0,0)$ is allowed for preparations if the two bases have one vector in common.  For measurement uncertainty, the point $(0,0)$ is attainable if the two observables are exactly jointly measurable, i.e.,  they commute. That corresponds to the much sharper condition that the two observables are the same up to a permutation of the outcomes.

It turns out that (at least in finite dimensional Hilbert spaces) the convex hull of the preparation uncertainty region contains the measurement uncertainty region for projective measurements \cite{SRW}. There are numerical counterexamples showing that the convex hull is needed here. It is also clear that such a statement cannot hold for general POVM measurements. Indeed if we add noise to some given measurements, preparation uncertainties go up, because all distributions become broader. However, measurement uncertainty goes down, typically to zero, because with sufficient noise it becomes possible to measure the given observables jointly.

\subsubsection{An Aside:\ Ozawa's error and disturbance}
Masanao Ozawa recently asserted that Heisenberg made a mistake regarding the relationship between measurement precision and disturbance.  The basis for this is a paper \cite{Oza} from 2003, in which he gives mathematical definitions of a root mean square error $\epsilon(Q)$ and a root mean square disturbance $\eta(P)$ referring to the microscope situation.  Ozawa then alleges that Heisenberg's claim in \cite{H} is that $\epsilon(Q)\eta(P)\geq\hbar/2$, and goes on to show that this relation is not generally true.  From this Ozawa concludes that Heisenberg was wrong.

It should be amply clear from the above reading of Heisenberg that he works entirely on a heuristic level, and never defined root mean square quantities or claimed any inequality. Turning one of his intuitions into a mathematical statement that can then be proved or disproved is an active process, that gets at least as much input from the person doing it as it gets from Heisenberg. When it comes to exact statements, Heisenberg by himself is generally not even wrong. So refuting Heisenberg at that level is simply not a worthy scientific target. The only interesting thing is to find out whether there is something to his intuitions after all. In the case of error and disturbance there clearly is, as we have demonstrated above. There may be other aspects, focusing on other quantities. But the one sure criterion that such an attempt has missed the target is to have no uncertainty relation. This is what happened to Ozawa.

Around the time of his paper, I (R.F.W.) heard Ozawa speak at a conference in Japan and tried immediately  to convince him that his formulation of measurement uncertainty and disturbance was no good.  I failed, but at least that stimulated me to show how one could do this better \cite{We1}.  One way or another, the subject hardly interested anyone for quite a while after that.  However, when an experimental group succeeded in measuring Ozawa's quantities \cite{RDMHSS}, huge hype in the media appeared stating that Heisenberg had been experimentally disproved.

Naturally, the experiment is quite irrelevant for the question of whether Heisenberg was correct.  It is also irrelevant for the question of whether Ozawa's definition of error and disturbance are sensible renditions of what one might understand by these terms. Measuring a silly quantity does not make it less silly.  The uncertainty relations that Ozawa foisted on Heisenberg do not even fail in any interesting way, as a quick look will show.  In fact, it was remarked in the literature long before Ozawa\cite{App} that one should not proceed in this way.  So from this point of view Ozawa's work was simply a step backwards.

One reaction to the hype of the supposed refutation of Heisenberg was that Paul Busch and Pekka Lahti got in touch with me (R.F.W.) and proposed that we should go against the hype on the basis of my old paper \cite{We1}.  Out of this came the stimulating collaborations \cite{BLW1,BLW2}, including a detailed criticism of Ozawa's approach \cite{BLW3}. 

\subsubsection{Uncertainty 10:\ No information gain without disturbance}
Till now, we have only considered measurement uncertainty for pairs of observables.  Sometimes it is interesting to allow arbitrarily many observables.  A situation where this becomes necessary is quantum cryptography, or, more precisely, quantum key distribution.  In this setting there is a channel shared by legitimate users, traditionally referred to as Alice and Bob, over which quantum particles are sent.  The eavesdropper is evil Eve.  She may carry out arbitrary measurements on the particles flying by.  But in so doing, she disturbs them, something Alice and Bob can readily determine by statistical tests.  The fundamental principle at work here is that there cannot be any information gain from a quantum system without disturbance. To make this explicit, consider a measurement process $\mathcal M$ with the property that the state after the measurement is always the same as before. That is, for any further statistical measurement it does not matter whether $\mathcal M$  was carried out or not. Then it follows that the results obtained by $\mathcal M$ are independent of the input state, i.e., we learn nothing about the quantum system.

Actually, we need a stronger statement, because exact equality of states before and after a transmission line never holds. There are always small losses, and the rules of the game are that we must attribute them to the eavesdropper. Thus we need to conclude from a sufficiently small disturbance, that the eavesdropper cannot have learned much, and give quantitative bounds for this. This is harder to get, and there are different ways of phrasing it mathematically. With an error criterion based on maximal errors one result of this kind is \cite{KSW}.  Moreover, entropic uncertainty relations with side information have proven to be useful for cryptography \cite{TR}.

It is interesting that the converse of the statement ``no information gain without disturbance'' also holds. It is known as the  Knill-Laflamme criterion for the correctability of errors \cite{KL}. If during some quantum process no information has been given to the environment, then there is a quantum process that restores the input state, i.e., there was no real disturbance. This is can be used to design fault-tolerant quantum computers, i.e., computers that work even with any desired accuracy in spite of imperfect components.

\subsection{Conclusions from later developments}
If ``knowledge'' about a quantum mechanical system is established via preparation or measurement, then we need preparation uncertainty relations and measurement uncertainty relations together to translate Heisenberg's idea into quantitative theorems in quantum mechanics.  The interest in doing this has grown since Heisenberg's time.  In his time, experiments close to the  uncertainty limit were hardly conceivable.  Today they are commonplace in laboratories.  There it is important to know how close exactly one is to the boundary.  In formulating such uncertainty relations there is considerable mathematical leeway, which one can use to construct and to prove inequalities tailored optimally to a given situation.  We should get used to speaking, not of ``the'' uncertainty relation, but rather of ``an'' uncertainty relation, or a whole collection of them.

Heisenberg demonstrated great foresight when he introduced his relations as a heuristic principle.  For many years afterwards, that was, without a doubt, their most important role, and still today there is nothing better to quickly decide where quantum effects must be taken into consideration.

\section*{Remarks}

\doenote{1} In a letter to Pauli \cite[{[105]}]{Meyenn}, Heisenberg laments the existence of two separate quantum communities in G\"ottingen, with very different views of the relationship of Mathematics and Physics. He especially dislikes the group around Hilbert and Weyl embracing matrices as bringing new progress in physics, and he even considers finding a more physicsy term for matrices. The two groups did make rather different contributions. Whereas Heisenberg was satisfied to develop an intuition ``for all simple cases'' that occurred to him, the mathematicians von Neumann and Weyl searched for and readily found the generalizable structures and interpretations that we still use today.  On the other hand, the uncertainty paper is an undeniable success of the heuristic physical, anti-mathematical approach.

However, some of the physicists around Heisenberg did feel the need to bring adequate mathematical tools into the new theory. In 1925/26 Born had already brushed up his operator theory on a visit to Norbert Wiener. He was also the Academy member to submit von Neumann's paper \cite{vN}. Just a short while later Pauli and Jordan were both collaborating with von Neumann.

Heisenberg in his later years expresses his appreciation of the mathematical side by claiming that it was his own work in the first place. Here is how, in 1956 [H5],  he summarizes the contribution of his uncertainty paper:\ the link from an experimental situation to its mathematical representation was to be achieved by the ``hypothesis that only such states may appear in nature, or can be realized experimentally, that can be represented by vectors in {\it Hilbert} space.'' (Italics in the original). Now this and much more could be said about von Neumann's article \cite{vN}, in which he actually coins the term ``Hilbert space''. Heisenberg's paper [H] naturally does not contain the word. Neither does it contain the thing itself.  In [H5] he also claims to have corresponded extensively with Pauli about ``this kind of solution'', but again, the surviving letters (e.g.\ \cite[{[115]}]{Meyenn}) have nothing.\por

\doenote{transl} We found two translations of \cite{H} (see references). Both make a mess of this distinction. Wheeler and Zurek choose ``physical content''. The anonymous translation on the NASA website has ``actual content''. Another option, found in the translation of a paper by Schr\"odinger, is  ``perspicuity''. See also \cite{HiUf}.\por

\doenote{matrix} In contrast to today, linear algebra was not part of the standard curriculum.  When he created matrix mechanics in 1925, Heisenberg knew nothing about the mathematics of matrices, and even Born found it worth mentioning from whom he himself had learned this exotic subject.\por

\doenote{2} The published statement, which Heisenberg presumably meant here, came from Schr\"odinger's paper \cite{Sch} in which he shows the equivalence of matrix mechanics and his theory. It is from a footnote in which he recalls why he had initially ignored Heisenberg's work and why, therefore, his own theory owed nothing to Heisenberg.  It is a pity that this is the only part of the paper that Heisenberg mentions. A more mature reaction would have been to accept Schr\"odinger's result that the two approaches are two sides of the same coin, with priority granted to matrix mechanics, and then work on the remaining differences.
The fight against the ``disgusting continuum theorists'' as a central issue of \cite{H} is worked out clearly in \cite{Bel}.\por

\doenote{3} ``We create internal virtual images or symbols of the external objects, and we make them in such a way that the logically necessary consequences of the images would invariably be the images of the natural consequences of the depicted objects.'' \cite{Hz}.  The role of the Hertzian ``images'' in Heisenberg's early works is traced in  \cite{Hen}.\por

\doenote{4} ``The philosophy is written in that great book, which constantly lies open before our eyes (I speak of the universe) \ellipsis.  It is written in the language of mathematics, and the letters are triangles, circles and other geometrical figures.''~\cite{Gal}\por

\doenote{ansch} In fact, the move hardly convinced Heisenberg himself. The ``victory footnote'' \citeH{196} cited above continues by granting Schr\"odinger an important role in the ``mathematical (and {\it in this sense} intuitive)'' development of the theory [italics by Heisenberg]. This supports my reading that the notion has been extended to include mathematical intuition, here even {\it any} mathematical work, but at the same time it seems visualization of a lesser kind. If we apply that to the paper \cite{H} itself it amounts to affirming the criticism it sets out to rebut:\
`Sorry, we do not have vizualizable content, but lots of math'. Therefore the visualizability grapes have to be sour. Indeed, Heisenberg continues by charging wave mechanics with ``leading away from the straight path outlined''  by Einstein, de Broglie, Bohr, and `{}quantum mechanics', by the poison of ``popular visualizability''. \por

\doenote{5} Nevertheless, the microscope yields further interesting aspects upon closer inspection.  For example, one can discuss what happens if one places the photographic plate or detector in the focal plane instead of the image plane, so that one detects the direction of the photon, not its point of origin.  Moreover, one can make this choice after the Compton scattering has occurred.~\cite{Her}\por

\doenote{13} An example is the online exhibition about Heisenberg by the American Institute of Physics \cite{onlineEX}.  Following the link ``Derivation of the uncertainty relation'', one finds the cited page. \por

\doenote{6} In the ``two-men-paper'' \cite{BJ}, this is the functional derivative of the formal trace of the Hamiltonian, which requires $H$ to be written as a non-commutative polynomial in a special symmetrized form.  In the ``three-man-paper'' with Heisenberg \cite{BHJ}, this is replaced by the directional derivative along scalar shifts.  That makes the transition to commutators and to the modern representation considerably simpler.  Both forms of partial derivative are forgotten today and only served the explicitly stated purpose of allowing the equations of motion to appear in a similar form to Hamilton's equations.\por

\doenote{7} The ``simplest conceivable assumption'' is actually a very poor, even wrong description of the transition. The distinction between diagonal and non-diagonal matrix elements depends on the basis (and none is specified), and just one diagonal element [=Diagonalglied] as a replacement is even crazier. Just a little further down the whole matrix is taken as the replacing object, which makes more sense. It is an interesting project to develop a reading of the quote that makes sense without excessive interpretational bias. In the paper the most helpful passage for this might be \citeH{181/182}. Like the quote, it seems to refer to a notion of assigning values to a matrix which is not wholly captured by an expectation value.
\por

\doenote{8} In \cite{H4} Heisenberg dates his conversation with Einstein, from which the quote presumably comes, to the spring of 1926.  He mentions that, for him, recalling that conversation was essentially the stimulus for the uncertainty paper.  In the narrative of \cite{H4}, Einstein  criticizes the methodology in \cite{H0} although Heisenberg feels that he has taken it from Einstein himself. It is hard to say whether Einstein would have found \cite{H} any more in agreement with his philosophy.\por

\doenote{9} The observable-centered approach of the older work \cite{H0} of 1925 is sufficient for Heisenberg here to proclaim the conclusiveness of the theory!  That any method for establishing a new theory guarantees its truth, must be doubted.  Moreover, this confidence is expressed towards a formalism that is still in its infancy.  To what extent the matrices of the formalism \cite{BHJ} are ``experimentally determined numbers'' remains unclear.  In \cite{BJ} and \cite{BHJ}, the concept-critical approach plays a subordinate role.  Hence, a purpose of this quoted sentence is to mention this aspect and with it Heisenberg's own contribution once again, and so to defend primacy and priority for the whole of quantum mechanics as well as its ultimate formalism (see also [*1]).\por

\doenote{10} If one reads them in isolation, one may exclaim at these sentences:\ ``Has this fellow not read Heisenberg?''.  Wasn't it the whole purpose of \S1 to cast doubt upon the word ``exact''?  But we hope it has become clear that ``exact'' in this sentence has nothing to do with the opposite of uncertainty or imprecision.  Rather ``exactly defined'' is to be read here as ``operationally defined''.\por

\doenote{11} How the uncertainties, which were introduced with the meagre precision of Heisenberg's tildes, could become the ``reason'' for precise quantitative probability relations is another unfulfilled
burden of proof. It only appears in the abstract. The paper itself has no details about it.\por

\doenote{14} Heisenberg uses a variant of this line also in his Nobel lecture \cite{Hnob}. The error of taking the uncertainty relation as an additional element, which is independent enough of quantum mechanics so that one could even think of taking or leaving it, has also been committed by some people without such personal motives.  For example, following his semi-classical discussion of the double-slit experiment, Richard Feynman writes:\ ``The uncertainty principle `protects' quantum mechanics.  Heisenberg recognized that if it were possible to measure the momentum and the position simultaneously with a greater accuracy, the quantum mechanics would collapse. So he proposed that it must be impossible. Then people sat down and tried to figure out ways of doing it, and nobody could \ellipsis. Quantum mechanics maintains its perilous but still correct existence.'' \cite[Section 1-8]{Fey}  That completely warps the role of a theorem in the theory and is also historical nonsense, just like Feynman's fictitious reasons for why Einstein did not accept quantum mechanics \cite[18-8]{Fey}.\por

\doenote{15} Don Howard has an entirely different interpretation here \cite{How}.  He would possibly let this be true for Heisenberg but is of the opinion that Bohr had, from the beginning, a deep understanding of entanglement, particularly the entanglement between a system and measuring apparatus caused by the measurement interaction.\por

\doenote{16} It is also unclear for me (R.F.W.) how the American Kennard came to write in a German style so reminiscent of Heisenberg's.  Maybe Heisenberg's contribution to this paper was bigger than Kennard's acknowledgement reveals. In a letter to Pauli \cite[{[164]}]{Meyenn}, dated May~31 in Copenhagen, Heisenberg writes:\ ``I am still very unhappy about the work of an American, which he began with me, and which is in that touchy subject, which I would like to stay away from right now''. This could be a reference to Kennard, but the normally very thorough editors of the letters give no hint about his identity. \por

\doenote{17}  A relatively early acknowledgement of this problem came from Karl Popper \cite{Pop1,Pop2}.  In these papers we see him fighting the ambiguity in \cite{H}.  He did not achieve a complete clarification, and later even had to retract a proposed experiment.  But the distinction between the ``statistical scatter relations'' (Poppers own translation of ``Streuungsrelationen'', i.e., variance relations) of preparation uncertainty from questions of measurement precision is quite clear:\ the measurement of a distribution presupposes the sharp measurability of the outcomes in each single case.  He comes very close to the concepts of preparation and measurement uncertainty with his distinction of non-prognostic measurements, i.e., those in which one does not care about the particle afterwards, and prognostic measurements, which serve to prepare a new initial state. In his immediate response, von Weizs\"acker gives an argument why the measurement uncertainty relations (which Popper is in some sense asking for) must be the same as the preparation uncertainty relations:\ he sees one kind directed towards the future and the other towards the past, and invokes the time inversion symmetry of quantum mechanics. If it were really that simple, we could have saved ourselves a lot of mathematical work. \por

\doenote{povm} This will usually be a generalized observable, for which the probability of each outcome is not given by a projection, but by a general positive operator (POVM, positive operator valued measure).  In a sequential measurement, this is anyhow what comes out, so we cannot avoid this kind of observable. However, it is a point that von Neumann missed. And here we have to say that great men cast long shadows. Even though his role in the early days of quantum mechanics is systematically ignored in the textbook literature on quantum theory, and most physicists have not even heard of him, von Neumann was in the long run hugely influential. His 1932 book was widely recognized as the mathematical basis, and his choice of projections as the yes/no observables, the related identification of observables with hermitian operators, and finally the projection postulate are almost universally accepted. However, these assumptions are unnecessarily restrictive. They are often false in experiments, and almost always when one analyzes an indirect measurement. Various authors came to this realization, beginning with Holevo in estimation theory and Ludwig \cite{Lud} for reasons of axiomatic parsimony. Now the quantum information community operates entirely in the wider framework. It would be interesting to study von Neumann's reasons, but that is beyond the scope of this article.\por

\end{document}